\documentclass[%
 reprint,
 amsmath,amssymb,
 aps,
]{revtex4-1}
\usepackage{url}
\usepackage[colorlinks,linkcolor=blue]{hyperref}
\usepackage{graphicx}
\usepackage{dcolumn}
\usepackage{bm}

\begin{document}
\title{Large-scale GHZ states through topologically protected  zero-energy mode in a superconducting qutrit-resonator chain}
\author{Jin-Xuan Han$^{1}$}\author{Jin-Lei Wu$^{1}$}\email[]{jinlei\_wu@126.com}
\author{Yan Wang$^{1}$}\author{Yan Xia$^{2}$}\author{Yong-Yuan Jiang$^{1}$}\author{Jie Song$^{1,3,4,5}$}\email[]{jsong@hit.edu.cn}
\affiliation{$^{1}$School of Physics, Harbin Institute of Technology, Harbin 150001, China\\
$^{2}$Department of Physics, Fuzhou University, Fuzhou 350002, China\\
$^{3}$Key Laboratory of Micro-Nano Optoelectronic Information System, Ministry of Industry and Information Technology, Harbin 150001, China\\
$^{4}$Key Laboratory of Micro-Optics and Photonic Technology of Heilongjiang Province, Harbin Institute of Technology, Harbin 150001, China\\
$^{5}$Collaborative Innovation Center of Extreme Optics, Shanxi University, Taiyuan, Shanxi 030006, People's Republic of China}

\begin{abstract}
We propose a superconducting qutrit-resonator chain model, and analytically work out forms of its topological edge states. The existence of the zero-energy mode enables to generate a state transfer between two ends of the chain, accompanied with state flips of all intermediate qutrits, based on which $N$-body Greenberger-Horne-Zeilinger~(GHZ) states can be generated with great robustness against disorders of coupling strengths. Three schemes of generating large-scale GHZ states are designed, each of which possesses the robustness against loss of qutrits or of resonators, meeting a certain performance requirement of different experimental devices. With experimentally feasible qutrit-resonator coupling strengths and available coherence times of qutrits and resonators, it has a potential to generate large-scale GHZ states among dozens of qutrits with a high fidelity. Further, we show the experimental consideration of generating GHZ states based on the circuit QED system, and discuss the prospect of realizing fast GHZ states.
\end{abstract}
\maketitle

\section{Introduction}
Greenberger-Horne-Zeilinger~(GHZ) states constitute an important class of entangled many-body states. A general form of $n$-qubit GHZ states can be expressed as~\cite{Bell1990,Bayesian2020,Reiter2016,Omran570,One2001}
\begin{equation}\label{e000}
\frac{1}{\sqrt{2}}(|\alpha_1 \alpha_2 \alpha_3\cdots \alpha_n \rangle+e^{i\phi}|\beta_1 \beta_2 \beta_3 \cdots \beta_n \rangle).
\end{equation}
where $\alpha_j+\beta_j=1~(\alpha_j,\beta_j\in\{0,1\})$. These states play a key role in quantum science and technologies, including open-destination quantum teleportation~\cite{Yang2004}, concatenated error-correcting codes~\cite{Knill2005}, quantum simulation~\cite{Song2018}, and high-precision spectroscopy~\cite{Leibfried2004}. In principle, the benchmark for quantum information capability is the number of particles that can be reliably entangled in a quantum processor. In experiment, multi-body entanglement is achieved recently by capturing 20 trapped ions with around the fidelity of $63.2\%$~\cite{Friis2018}, 12 photons with $59.8\%$~\cite{Zhong2018}, 18 photonic qubits exploiting three degrees of freedom of six photons with $72.4\%$~\cite{Wang2018}, 12 superconducting qubits with $55.6\%$~\cite{Gong2019}, 18 superconducting qubits with $53.0\%$~\cite{Song2019} and 20 Rydberg atoms with $54.2\%$~\cite{Omran2019}. Among various physical systems, the ubiquitous noise and device imperfections are, however, unavoidable and limit the range where  multi-body entanglement can be realized with a high fidelity. 

Topological insulator~\cite{Qi2011,Chiu2016}, a new kind of novel state
of matters, is characterized by the conducting edge states and the insulating bulk states. The
special conducting edge states are protected by the energy gap of the topological system, leading the edge state to be insensitive to local perturbations and disorders \cite{Kitaev2001,Kane2005,Hasan2010}. To this end, the topological state of matter has been studied in many systems, such as optical lattice systems~\cite{Guo2016,Lin2015}, spin Fermi systems~\cite{Ren2015,Shih2016}, supercooled atoms~\cite{Jotzu2014,Aidelsburger2015} and synthetic materials~\cite{Sun2017}. These novel properties support many potential applications of topological insulators in quantum information processing and computing. For instance, several proposals for quantum state transfer (QST) have been presented ~\cite{Yao2013,Dlaska2017,Mei2018,Longhi2019,Qi2020,Tan2020,Zheng2020,Qi2020pra} with topologically protected channels. A topologically protected channel for QST between remote quantum nodes mediated by the edge mode of a chiral spin liquid was proposed and analyzed~\cite{Yao2013,Dlaska2017}. Mei $et~al.$ presented an experimentally feasible mechanism for implementing robust QST via the topological edge states by connecting superconducting Xmon qubits into a one-dimensional chain~\cite{Mei2018}. Also, topologically protected entangled photonic states and its transport via edge states have been reported~\cite{Mikael2016,Blanco2018,Michelle2019,Hu2020}. In experiment, the topological protection of spatially entangled biphoton states was demonstrated~\cite{Michelle2019}. The robustness in crucial features of the topological biphoton correlation map in the presence of deliberately introduced disorder was found in the silicon nanophotonic structure. Recently, a spatially entangled two-particle NOON state is proposed by topological Thouless pumping in one-dimensional disordered lattices~\cite{Hu2020}.

In this paper, we propose a superconducting-circuit model to generate large-scale GHZ states, where the model is a chain consisting of $N$ flux qutrits connected by $(N-1)$ resonators, and the entangled states are protected by topological zero-energy mode. We analytically derive the wave function of the topological edge state with zero energy of the qutrit-resonator chain, through which a state transfer between two ends of the chain, accompanied with the state flips of all intermediate qutrits, can be implemented by designing qutrit-resonator coupling strengths. The key advantage that topology offers in such processes is the inherent protection of boundary edge states lying in the band gap of the dispersion relation when the bulk is topologically nontrivial. Based on such a peculiar state transfer protected topologically, we show three schemes of generating large-scale GHZ states, providing feasible and visible methods to generate the robust large-scale GHZ states by meeting the performance requirements of different experimental devices in the superconducting qutrit-resonator chain. Meanwhile, we take into account the experimental consideration, such as the implementation of the model, preparation of the initial state, and realization of tunable couplings in the circuit-QED chain. In addition, we discuss the potential of realizing fast GHZ states by speeding up the adiabatic proscess, which provides more possibilities for obtaining high-fidelity multi-body entanglement.

Our work may facilitate the potential applications of topological matter in quantum information processing, due to the following advantages and interests. First, the topological qutrit-resonator chain can be used to create a large-scale GHZ state theoretically with the size being far more than $N=20$ that is the particle number of multi-body entanglement realized experimentally up to now~\cite{Friis2018,Zhong2018,Wang2018,Gong2019,Song2019,Omran2019}. Second, as the core foundation of implementing large-scale GHZ states, we derive theoretically the wave function of an edge state with zero-energy mode, whose form involves the state flips of all intermediate qutrits, different from that in the frequently studied standard Su-Schrieffer-Heeger (SSH) model~\cite{Mei2018,Qi2020,Tan2020,Zheng2020}.
Finally, there are three schemes proposed for generating large-scale GHZ states, which provide potential choices, depending on different device requirements, i.e., coherence times of qutrits and resonators.

\section{Physical model and wave function of an edge state}\label{S2}
\begin{figure*}\centering
\includegraphics[width=0.9\linewidth]{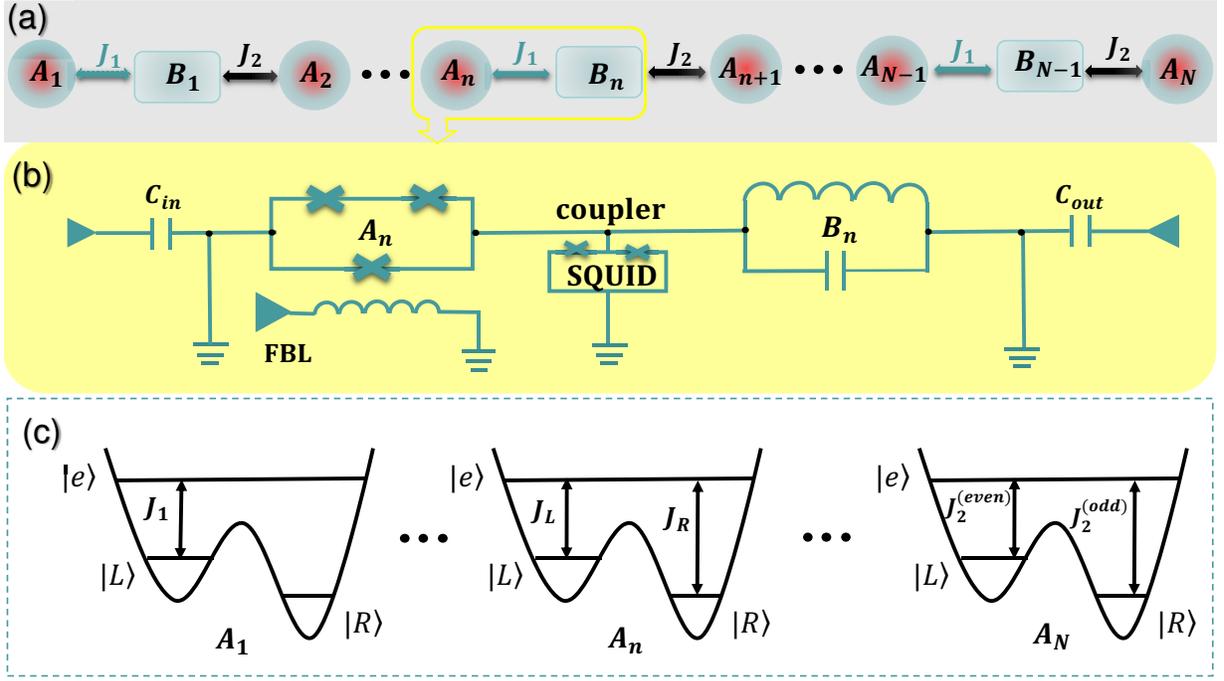}
\caption{(a)~A superconducting qutrit-resonator chain. The $n$-th unit cell contains one flux qutrit and one single-mode resonator, labeled as $A_n$ and $B_n$, respectively, and holds an intra-cell qutrit-resonator coupling strength $J_1$. Between two adjacent cells, a qutrit $A_{n+1}$ is coupled to the resonator $B_n$ with an inter-cell coupling strength $J_2$. The resonator $B_n$ drives the transition $|L\rangle \leftrightarrow |e\rangle$ ($|R\rangle \leftrightarrow |e\rangle$) of the two nearest-neighbor qutrits $A_n$ and $A_{n+1}$ when $n$ is odd (even). (b)~Circuit schematic of one unit cell in the superconducting qutrit-resonator chain. The coupling strength can be dynamically tuned by a coupler of SQUID. The energy level space of the qutrit can be tuned by FBL. (c)~Schematics of energy level transitions for qutrits $A_1$, $A_n$~($1<n<N$), and $A_N$. The energy level structure of a flux qutrit holds two ground states ($|L\rangle$ and $|R\rangle$) and one excited state ($|e\rangle$). We denote that coupling strengths $J_L=J_1$ and $J_R=J_2$~($J_L=J_2$ and $J_R=J_1$) when $n$ is even (odd).}\label{f1}
\end{figure*}

\subsection{Physical model}\label{S2A}
The setup of the superconducting qutrit-resonator chain for generating large-scale GHZ states is illustrated in Fig.~\ref{f1}(a). The chain contains $(2N-1)$ lattice sites, $N$ qutrits and $(N-1)$ resonators. Each unit cell in the chain contains one flux qutrit $A_n$ and one resonator $B_n$, whose circuit schematic is described by Fig.~\ref{f1}(b). The energy levels of qutrit and the coupling strengths are adjustable via the magnetic flux provided by the flux-bias line~(FBL)~\cite{Schmidt2013} and superconducting quantum interference device (SQUID). As described in Fig.~\ref{f1}(c), each flux qutirt holds a three-level structure, involving two ground states $|L\rangle$ and $|R\rangle$, and one excited state $|e\rangle$.
The interaction in the chain can be described by the following interaction-picture Hamiltonian ($\hbar$ = 1)
\begin{equation}\label{e0}
H_{I}=\sum_{n=1}^{N-1}\big(J_1|e\rangle_{n} \langle j_n|+ J_2|e\rangle _{n+1}\langle j_n|\big)b_n+\rm{H.c.},
\end{equation}
where $j_n=L~(R)$ when $n$ is odd~(even), and $b_{{n}}$ the annihilation operator of the resonator $B_n$. $J_1$ and $J_2$ can be tuned through adopting controlled voltage pulses generated by an arbitrary waveform generator (AWG) to tune the flux threading the loop~\cite{Majer2009}. Further, $H_{I}$ can be rewritten as
\begin{equation}\label{e1}
H=\sum_{n=1}^{N-1}J_1b_n\sigma_{n}^+(\sigma_{n}^{x})^{n}+J_2b_{n}\sigma_{n+1}^+(\sigma_{n+1}^{x})^{n}+\rm{H.c.},
\end{equation}
where we define $\sigma_{n}^+=|e\rangle_n\langle R|$, $\sigma_{n}^{x}=|L\rangle_n\langle R|+|R\rangle_n\langle L|$. The existence of the $(\sigma_{n}^{x})^n$ renders two different transitions $|L\rangle  \leftrightarrow |e\rangle$ and $|R\rangle  \leftrightarrow |e\rangle$ for qutrit $A_n$ depending on the odevity of $n$. For instance, $\sigma_{n}^+(\sigma_{n}^{x})^{n}=|e\rangle_n \langle R|~(|e\rangle_n \langle L|)$ with $n$ being even~(odd).

Such a adjustable chain can be analogous to an SSH model~\cite{Soliton1980,short2016,Su1979} which describes quanta~(e.g., electrons, photons, or phonons) hopping on a chain (one-dimensional lattice), with staggered hopping amplitudes. The chain of SSH model consists of $N$ unit cells, each unit cell hosting two sites, one on sublattice $A$, and the other on sublattice $B$. The Hamiltonian of the standard SSH model is of the form $H_{SSH}=(\nu\sum_{m=1}^{N}|m,B\rangle\langle m,A|+\mu\sum_{m=1}^{N-1}|m+1,A\rangle\langle m,B|)+\rm{H.c.}$. Here $|m,A\rangle$ and $|m,B\rangle$, with $m \in {1,2,\cdots N}$, denote the states of the chain where the hopping quantum is on sites $A$ and $B$, respectively, in the unit cell $m$~\cite{Soliton1980,short2016}. The Hamiltonian~(\ref{e1}) holds a form of the SSH model whose topological phase is characterized by winding number~\cite{Ryu_2010}, except for an additional operator $(\sigma_{n}^{x})^{n}$ that works for flipping qutrit states and is the key to realizing GHZ states. Based on the bulk-edge correspondence~\cite{Qi2010,Hasan2010,Classification2016}, the SSH model possesses zero-energy edge modes at open boundaries in the topologically nontrivial phase, which are protected by the topological properties of the system.

\subsection{Wave function of an edge state}\label{S2B}
The appearance of topologically protected gapless edge states within the bulk gap is a manifestation of the topological insulator. The number of such gapless edge modes is specified by topological invariants. As for an SSH model, the zero energy mode, one of the characteristics of the topological nontrivial SSH phase, is regarded as the topological invariant in the edge state and will protect the edge state topologically from local disorders. Thus, it is critical to obtain the wave function of the edge state with zero energy mode in our model.

The edge states of the chain with a single excitation are exponentially localized at the boundaries. The wave function of an edge state can be described by the following ansatz, analogous to the standard SSH model~\cite{Soliton1980,short2016,Su1979}
\begin{equation}\label{e2}
|\varphi_e\rangle=\sum_{n=1}^{N}\lambda^{n}\Big[\gamma\sigma_{n}^+(\sigma_{n}^{x})^{n-1}\bigotimes^{n-1}_{l=1}\sigma_{l}^{x}+\eta b_{{n}}^{\dagger}\bigotimes^{n}_{l=1}\sigma_{l}^{x}\Big]|G\rangle,
\end{equation}
where
\begin{equation*}
|G\rangle=|\underbrace{RLR\cdots}_{N} \rangle_{A_{N}}\otimes |\underbrace{000\cdots}_{N-1}\rangle_{B_{N-1}}
\end{equation*}
denotes a decoupled state of the qutrit-resonator chain with all resonators in $|0\rangle$~(i.e., zero-photon Fock state), while qutrits $A_1$, $A_2$, $A_3$, $\cdots$ in states $|R\rangle$, $|L\rangle$, $|R\rangle$, $\cdots$, repsectively. $\lambda$ is the localized index, $\gamma$ and $\eta$ being the probability amplitudes of the gap states. When $\lambda \textless 1$ ($\lambda \textgreater 1$), the probability amplitude of the site $n$ decays (increases) exponentially with distance $n$ which means the left (right) edge state with the wave function normalized. When $\gamma=0$~($\eta=0$), according to Eq.~(\ref{e2}) the resonators~(qutrits) are occupied by the edge state whose eigenenergy is $E=0$. In particular, in order to generate large-scale GHZ states with binary quantum information carried by two states in qutrits, we choose $\gamma=1$ and $\eta=0$ to render the qutrit in each unit cell to occupy the left~($\lambda<1$) and right~($\lambda>1$) edge states with $E=0$. Thus, the wave function of an edge state occupied by qutrits can be written as $|\varphi'_e\rangle=\sum_{n=1}^{N}\lambda^{n}\sigma_{n}^+(\sigma_{n}^{x})^{n-1}\bigotimes^{n-1}_{l=1}\sigma_{{l}}^{x}|G\rangle$. Substituting $|\varphi_e'\rangle$ into the eigenvalue equation $E|\varphi_e'\rangle=H|\varphi_e'\rangle$, one can obtain
\begin{eqnarray}\label{e3}
&&E\sum_{n=1}^{N}\lambda^{n}\sigma_{n}^+(\sigma_{n}^{x})^{n-1}\bigotimes^{n-1}_{l=1}\sigma_{{l}}^{x}|G\rangle\nonumber\\
&&=\Big(J_1\lambda^{n} b_{{n}}^{\dagger}\bigotimes_{l=1}^{n}\sigma_{{l}}^{x}
+J_2\lambda^{n+1}b_n^{\dagger}\bigotimes_{l=1}^{n}\sigma_{{l}}^{x}\Big)|G\rangle.
\end{eqnarray}
Through these analyses, the edge state can be worked out as
\begin{equation}\label{e4}
|\varphi_E\rangle=\sum_{n=1}^{N}\lambda^{n}\sigma^+_{{n}}(\sigma_{n}^{x})^{n-1}\bigotimes^{n-1}_{l=1}\sigma_{{l}}^{x}|G\rangle,
\end{equation}
with $\lambda=-{J_1}/{J_2}$.

\begin{figure*}
\centering
\includegraphics[width=0.75\linewidth]{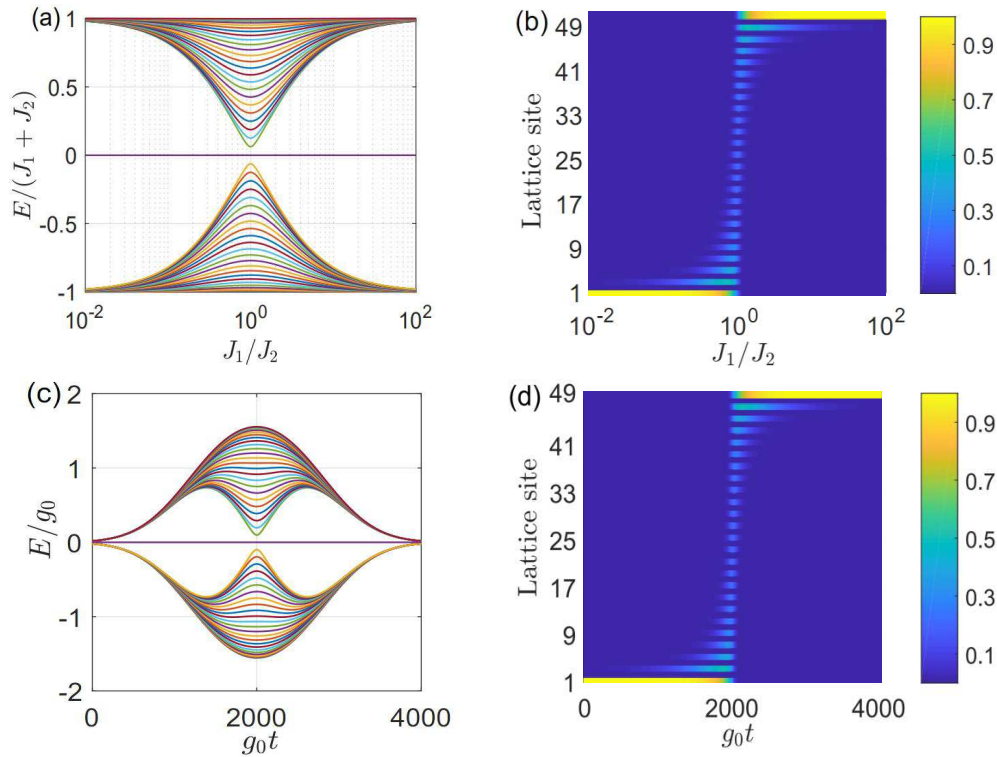}	
\caption{(a) Energy spectra of the chain versus $J_1/J_2$. (b) Distribution of the zero energy mode versus $J_1/J_2$. (c) Energy spectra of the chain versus $g_0t$. (d) Distribution of the zero energy mode versus $g_0t$. The size of the chain is $L=2N-1=49$.}\label{f2}
\end{figure*}

To illustrate the topological property of the qutrit-resonator chain more clearly, we take the size $L\equiv2N-1=49$ of the chain as an example to plot the energy spectrum of the system and the distribution of the topological edge states, respectively, in Figs.~\ref{f2}(a) and (b) with varying $J_1/J_2$. We find that the energy spectrum possesses a unique zero energy mode keeping unchanged with varying $J_1/J_2$, which denotes a non-evolutive state. The closest distance between the zero energy state and the bulk appears at the point $J_1/J_2=1$. In Fig.~\ref{f2}(b), the zero energy state is localized near the left extremity when $J_1/J_2 <1$, while for $J_1/J_2>1$ it is localized near the right extremity. It means that one can achieve the topological state transfer assisted by the zero energy mode between the first and the last qutrits via varying the parameter $J_1/J_2$, which is consistent with conclusion obtained from Eq.~(\ref{e4}). Therefore, the Hamiltonian $H$ is a  modulated model if varying the coupling strength $J_1$ and $J_2$. Specifically, the shapes of the coupling strengths are engineered as Gauss functions
\begin{equation}\label{e5}
\begin{split}
J_1&=g_0\exp{[-(t-3\tau)^2/\tau^2]},\\
J_2&=g_0\exp{[-(t-2\tau)^2/\tau^2]},
\end{split}
\end{equation}
where $\tau=T/7$ is set as the value of the pulse width, as well as the delay of $J_1$ and $J_2$, with $T$ being the evolution time. The forms of $J_1$ and $J_2$ in Eq.~(\ref{e5}) satisfy the state transfer conditions $J_1/J_2|_{t\rightarrow0}=0$ and $J_1/J_2|_{t\rightarrow T}=+\infty$ well. The idea of engineering the coupling strengths as Gauss functions is to achieve a temporal soft quantum control starting from and ending at a zero amplitude, which enables on-resonant couplings among a desired set of target levels, while efficiently avoiding unwanted off-resonant contributions coming
from others~\cite{Haase2018,Wu2020}.

Based on time-dependent Gaussian coupling strengths, as shown in Fig.~\ref{f2}(c), there exists a zero energy mode among all eigenstates during the whole evolutionary process. Simultaneously, the topological edge state at zero energy is well separated from bulk states. In Fig.~\ref{f2}(d), we plot the state distribution of the zero mode. The qutrit-resonator chain has not only a bulk part but also boundaries (which we refer to ends or edges). The qutrits $A_1$ and $A_N$ are regarded as two edges, while the other qutrits and resonators are the bulk part. The distribution of the zero mode state decays exponentially on the qutrits under the condition of $g_0t<1800$. Particularly, the eigenstate of the zero energy is localized near the first lattice site $A_1$~(left edge) and the distribution of left edge state with zero mode is equal to unity when $g_0t<1800$. When $g_0t$ increases continuously, the distribution of zero mode increases exponentially on the qutrits. Similarly, the eigenstate is localized near the last lattice site $A_{25}$ (right edge) and the distribution of right edge state with zero mode is equal to unity when $g_0t>2200$.

\section{Large-scale GHZ states}\label{S3}
\subsection{Scheme A for generating large-scale GHZ states}\label{S3A}
Now we focus on the generation of large-scale GHZ states among the $N$ qutrits in the superconducting qutrit-resonator chain. In Fig.~\ref{f2}(d), the edge state concentrates towards the left (right) end when $g_0t<1800$~($g_0t>2200$). In particular, when $g_0t\rightarrow 0$ and $g_0t\rightarrow +\infty$, the edge states become
\begin{eqnarray}\label{e01}	
|l\rangle&=&|eLR\cdots m\rangle_{A_{N}}\otimes|000\cdots0\rangle_{B_{N-1}},\nonumber\\
|r\rangle&=&|LRL\cdots e\rangle_{A_{N}}\otimes|000\cdots0\rangle_{B_{N-1}},
\end{eqnarray}
where $m=R~(L)$ when $N$ is odd (even). For generating GHZ states, the superconducting qutrit-resonator chain is assumed initially in the state
\begin{eqnarray}\label{e6}
|\phi_0\rangle&=&\frac{1}{\sqrt{2}}\left(|G\rangle+|l\rangle\right).
\end{eqnarray}
Going through the evolution, the first decoupled state component in the Eq.~(\ref{e6}) does not evolve because no photon in the resonators can be absorbed to excite the ground-state qutrits. From the Fig.~\ref{f2}(d), we learn that the second term of Eq.~(\ref{e6}) is essentially the left edge state with zero energy, which can evolve into the right edge state along the topologically protected process. Thus, the following evolution occurs
\begin{equation*}
|\phi_0\rangle~\mapsto~\frac{1}{\sqrt{2}}\left[|G\rangle-(-1)^{N}|r\rangle\right].
\end{equation*}
Here we set that $|R\rangle$ is used to carry the logical state 1 and $|L\rangle$ is used to carry the logical state 0 for the qutrit $A_n$~($1\leq n<N$), while $|e\rangle$ of the last qutrit $A_N$ is encoded as the logic state 1(0) when $N$ is even (odd). Except for the last qutrit $A_N$, the excitation state $|e\rangle$ is an auxiliary state without carrying any quantum information. Accordingly, the final state of $N$ qutrits after the evolution with omitting the zero-photon product state of resonators is 
\begin{equation}\label{e7}
\frac{1}{\sqrt{2}}(|101\cdots 1(0)\rangle_{A_{N}}-(-1)^{N}|010 \cdots 0(1)\rangle_{A_{N}}),
\end{equation}
which is exactly an $N$-body GHZ state according to Eq.~(\ref{e000}).

\begin{figure}[htb]\centering
\centering
\includegraphics[width=1\linewidth]{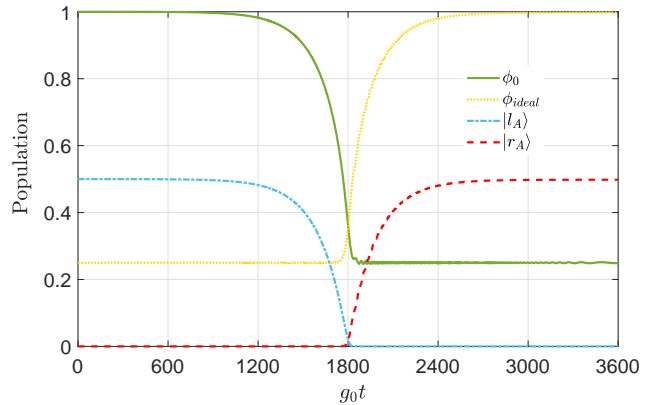}
\caption{Time evolution of populations for $|\phi_0\rangle$, $|\phi_{ideal}\rangle$, $|l_{\rm A}\rangle$ and $|r_{\rm A}\rangle$ with the size of the superconducting qutrit-resonator chain $L=49$. We choose $g_0T=3600$ as a total evolution time.}\label{f3}
\end{figure}

As an example, with the full Hamiltonian Eq.~(\ref{e1}) we take $N=25$ to numerically plot in Fig.~\ref{f3} the population evolution of the initial state, the ideal state $|\phi_{ideal}\rangle=\frac{1}{\sqrt{2}}(|RLR\cdots R\rangle_{A_{25}}+|LRL \cdots e\rangle_{A_{25}})\otimes|000\cdots 0\rangle_{B_{24}}$, and two edge states $|l_{\rm A}\rangle=|eLR\cdots R\rangle_{A_{25}}\otimes|000\cdots0\rangle_{B_{24}}$ and $|r_{\rm A}\rangle=|LRL\cdots e\rangle_{A_{25}}\otimes|000\cdots0\rangle_{B_{24}}$.
Evidently, the population of $|\phi_{ideal}\rangle$ evolves from $0.25$ to $1$, which indicates the successful creation of a $25$-body GHZ state. When $g_0t\approx 1800$, the populations of $|\phi_{ideal}\rangle$ and $|r_{\rm A}\rangle$ suddenly increase, which shows the same result as the distribution of the zero mode in Fig.~\ref{f2}(d).

\begin{figure*}[htb]\centering
\centering
\includegraphics[width=0.75\linewidth]{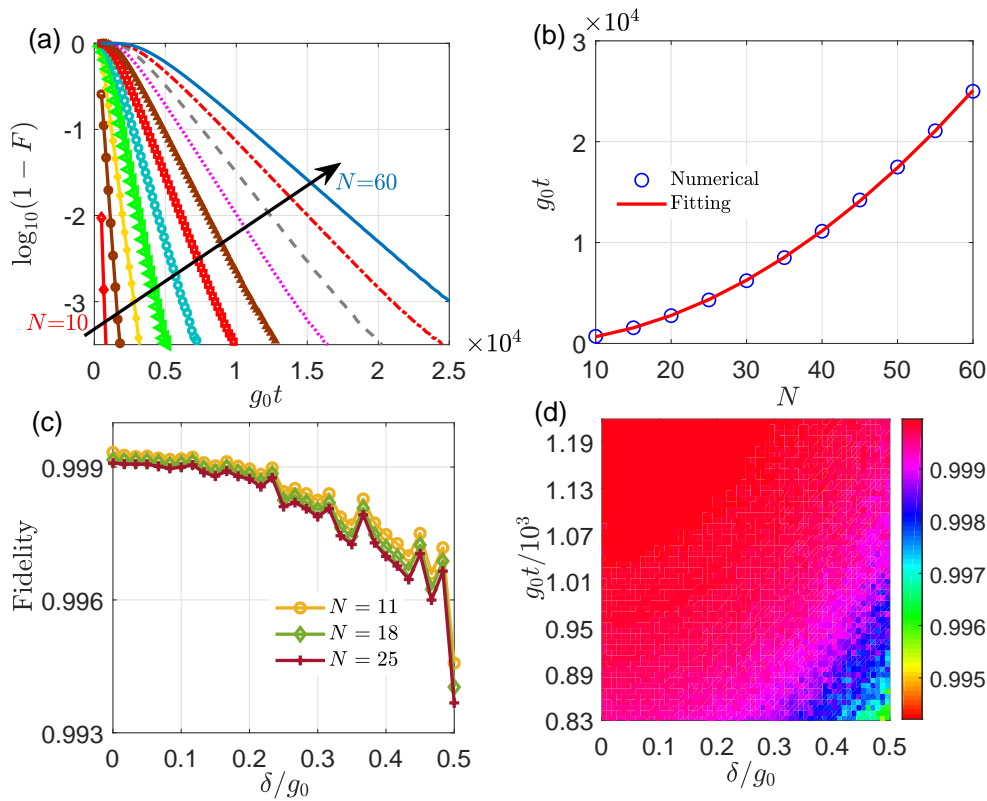}
\caption{(a)~Time evolution of $\log_{10}(1-F)$ for generating $N$-body GHZ states with $N$ from 10 to 60 at intervals of 5. (b) The fitting function and numerical scatters between the number of qutrits and evolution time with $99.9\%$ fidelity. (c) The fidelity of GHZ state against the coupling strength with disorder $\delta$ corresponding to 11, 18 and 25 qutrits. (d) The fidelity of GHZ state versus the varying $\delta/g_0$ and $g_0t/10^{3}$ for $11$-body GHZ state.}\label{f4}
\end{figure*}

\subsection{Shortest time for generating high-fidelity $N$-body GHZ states and robustness against disorder of coupling strengths}
In the protocol above, the $N$-body GHZ state is achieved through an excitation displacement of the zero-energy mode along the chain. The evolution time $T$ must be chosen as long as possible in order to obtain an adiabatic excitation displacement from one extremity to the other of the qutrit-resonator chain. The adiabaticity condition can be written as $\dot{\theta} \ll \Delta E$, where $\theta=\arctan(J_1/J_2)$ and $\Delta E$ is the energy gap between the zero-energy mode and the bulk modes~\cite{Mei2018,Felippo2020}. In the above numerical simulation, the evolution time is appropriate but not the shortest time of generating a high-fidelity~($0.999$) ideal GHZ state. 

In order to get the information of the shortest time of generating $N$-body GHZ states, we numerically calculate the fidelity of ideal $N$-body GHZ states, and plot in Fig.~\ref{f4}(a) the time evolution of $\log_{10}(1-F)$ with $N$ from 10 to 60 at intervals of 5. The numerical results exhibit that a higher fidelity needs a longer evolution time with increasing $N$. For example, the evolution time for $10$-body GHZ state is $g_0t=661$ with $99.9\%$ fidelity. However, $60$-body GHZ state with $99.9\%$ fidelity requires $g_0t = 2.5\times10^4$. In practice, the value of $g_0$ can be chosen about $2\pi\times$ $50$~MHz~\cite{Mundada2019}. Thus, the evolution time of realizing $60$-body GHZ state is $T \approx 76~\mu $s. Up to now, the superconducting resonator lifetimes can be achieved between $1$ and $10$ ms~\cite{Reagor2013,Reagor2016,Axline2016}. And the decoherence time of the superconducting magnetic flux qubtrit achieved on $1$ ms by designing a $\pi$-phase difference across the Josephson junction in circuit has been reported~\cite{Pop2014}, which is far more than the evolution time $T$. As shown in Fig.~\ref{f4}(b), by selecting different $N$ and corresponding evolution times of generating GHZ states with the fidelity of $99.9\%$ as numerical samples, $g_0t$ versus $N$ can be fitted by a quadratic function $g_0t=6.9419N^{2}+2.455N-59.8933$. The evolution time should be larger enough to satisfy the adiabatic condition for a longer size of the chain.

Because of the existence of some circuit imperfections in a superconducting qutrit-resonator chain such as the mutual inductance and self-inductance of the circuit, and the unstable magnetic flux threading to the SQUID loop, which may cause variation in ideal couplings. In order to study the robustness of the scheme, we add a random disorder into coupling strength $J'_{1(2)}=J_{1(2)}(1+\rm{rand}~[-\delta,\delta])$ for each qutrit-resonator coupling, where $\rm{rand}~[-\delta, \delta]$ denotes a random number within the range of $[-\delta,\delta]$. Figure \ref{f4}(c) shows the relationship between the fidelity of GHZ state and the disorder $\delta \in [0,0.5]$ for 11, 18 and 25 qutrits. Note that the disorder is randomly sampled 101 times, and then the fidelity is taken as the mean value of the 101 results. Corresponding to a shorter size of the superconducting qutrit-resonator chain, the fidelity decreases more slowly and the GHZ state is more robust to the disorder of coupling strength. On account of the process of generating the GHZ state via the topological zero mode protected by the energy gap, the width of the gap in a superconducting qutrit-resonator chain usually exhibits an exponential decay behavior with the size of chain increasing~\cite{Chiu2016}. Furthermore, the disorder has little effect on the GHZ state with respect to $\delta/g_0 <[0,0.2]$. Even if $\delta/g_0 \in[0.2,0.5]$, the fidelity still keeps over $0.993$, which reflects the topological protection property of the system. As shown in Fig.~\ref{f4}(d), we plot the fidelity versus the varying evolution time and the disorder of coupling strength $\delta$ for 11 qutrits. We can learn that the damage to fidelity caused by coupling defects $\delta/g_0 \in[0.2,0.5]$ can be compensated by longer evolution time. Under the condition of $\{g_0t=830, \delta=0.5\}$, the fidelity of 11-body GHZ state can be achieved with fidelity over $0.995$. The robustness of $N$-body GHZ state to the disorder and perturbation provides much more convenience for the experimental realization and the practical application of multi-particle entanglement state.

\subsection{Influence of the losses in the superconducting qutrit-resonator chain}
We now give a discussion for the effect of qutrit-resonator losses on the fidelity of generating GHZ states. Two dominant channels are considered in the loss mechanism: (i) The loss of qutrits with decay rates $\gamma_{n}$; (ii) The loss of resonators with decay rates $\kappa_{n}$.

The effect of losses during the evolution time can be evaluated by using a conditional Hamiltonian~\cite{Pachos2002,Huang2018}
\begin{eqnarray}\label{e8}
H_{cond}&=&H-\frac{i\kappa_n}{2}\sum_{n=1}^{N-1}b^{\dagger}_{n}b_{n}-\frac{i\gamma_n}{2}\sum_{n=1}^{N}|e\rangle_{n}\langle e|,
\end{eqnarray}
where $H$ is the lossless Hamiltonian for the system in Eq.~(\ref{e1}). The second and third terms represent the losses of resonators and qutrits, respectively. For convenience, we assume that $\kappa_{n}=\kappa$ and $\gamma_{n}=\gamma$. We take $N=11$, 18, and 25 as examples. The fidelity is formulated as $F=\langle \Psi_{ideal}|\rho|\Psi_{ideal}\rangle$, where $|\Psi_{ideal}\rangle=1\ \sqrt{2}(|G\rangle-(-1)^{N}|r\rangle)$ is the output state of an ideal system, and $\rho$ is the density operator of the system dominated by the conditional Hamiltonian Eq.~(\ref{e8}). 

\begin{figure}
\includegraphics[width=\linewidth]{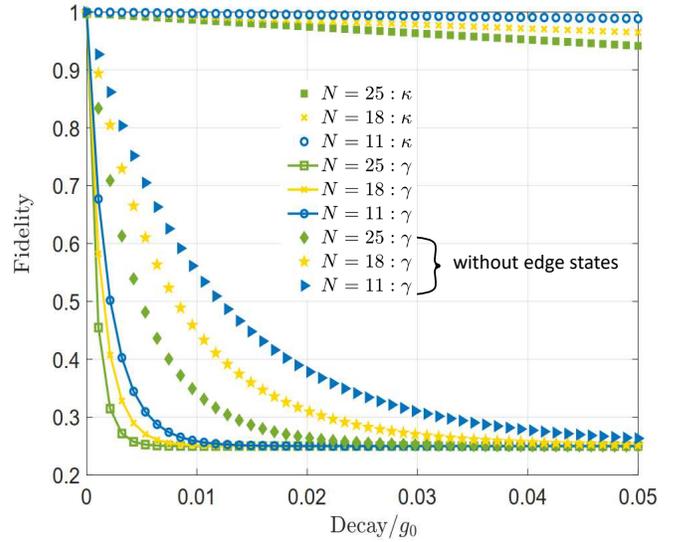}
\caption{Fidelities for GHZ states versus decay rates of qutrits or resonators for different $N$, where $\gamma$~($\kappa$) in the legend represents decoherence involving only the qutrit~(resonator) decay. The evolution time $T$ is chosen as $g_0T=6.9416N^{2}+2.455N-59.8933$ that ensures a $99.9\%$ fidelity for generating a lossless $N$-body GHZ state.}\label{f5}
\end{figure}

We now numerically simulate the fidelity of ideal GHZ states by solving the non-Hermitian Liouville equation $\dot{\rho}=-i(H_{cond} \rho-\rho H^{\dagger}_{cond})$. Figure \ref{f5} shows the relationship between the fidelity $F$ and decay of qutrits or resonators. As the decay rate of qutrits or resonators increases, the fidelity of the ideal state shows a trend of decline. Comparing those lines, the size of chain increases, the fidelity decreases slightly. And the higher fidelities indicate that the Scheme A for generating $N$-body GHZ states is insensitive to the value of $\kappa$. The fidelity of the 11-, 18- or 25-body GHZ state plummets to $25\%$ roughly with $\gamma/g_0=0.01$. Evidently, the value of $\gamma$ has a greater influence on the fidelity. Owing to the distribution of edge states $|l_{\rm A}\rangle$ and $|r_{\rm A}\rangle$ closed to 0.5 at end of the evolution time in Fig.~\ref{f2}(d), the loss of qutrits in the excited state $|e\rangle$ in edge states during the whole evolution time causes a significant destructive effect on the fidelity of GHZ states. As shown in Fig.~\ref{f5}, we plot the fidelity of GHZ states without considering the losses in left and right edge states, i.e, setting $\gamma_1=\gamma_n=0$. Under the condition of $\gamma_n/g_0=0.01$ for $1<n<N$, the fidelities of 11-, 18- and 25-body GHZ states are $55.24\%$, $44.72\%$ and $33.92\%$, respectively, which indicates a notable improvement of the fidelities than before. Consequently, how to suppress the loss of excited state $|e\rangle$ of qutrits, especially of $A_1$ and $A_N$,  is the key to enhancing the fidelity of GHZ states. In the next section, we propose two alternative schemes that are of improved robustness against losses of qutrits.

\begin{figure*}[htb]\centering
\centering
\includegraphics[width=0.7\linewidth]{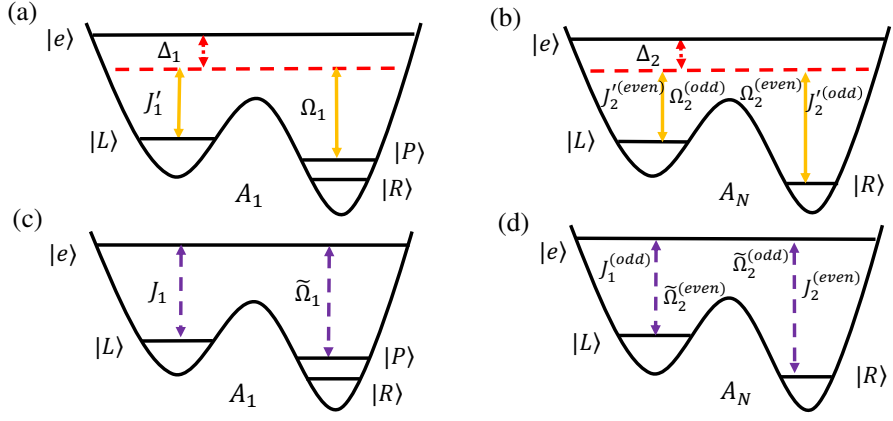}
\caption{(a) and (b): Schematics of energy level transitions for qutirts $A_1$ and $A_N$ in Scheme B. (c) and (d): Schematics of energy level transitions for qutirts $A_1$ and $A_N$ in Scheme C. The interaction diagram of other cells is the same as Fig.~\ref{f2} in both Scheme A and B. The choice of coupling strengths and driving fields of last qutrits $A_N$ depends on the odevity of $N$.}\label{f6}
\end{figure*}
\section{Alternative schemes}\label{S4}

In the process of generating GHZ states, the distribution of excitations in two edge states is 0.5 at the start and end of the evolution time~(see Fig.~\ref{f3}). The losses of qutrits in the excited states have a great destructive effect on the final fidelities of GHZ states. In this section, we propose two improvement protocols, labeled Scheme B and Scheme C, to focus on suppressing the loss of qutrits in the excited state.

\subsection{Scheme B for suppressing excitation of qutrits $A_1$ and $A_N$}
The energy level space of a qutrit can be readily adjusted via changing the external flux applied to the SQUID loop~\cite{Neeley2008,Yang2010,You2005} or/and provided by an FBL~\cite{Schmidt2013}. As shown in Figs.~\ref{f6}(a) and (b), an extra long-lived state $|P\rangle$ is introduced into the right potential well of the qutrit $A_1$, upper than and separated enough from the ground state $|R\rangle$. The transition between $|e\rangle \leftrightarrow |L\rangle$ in qutrit $A_1$ is coupled off-resonantly to the resonator $B_1$ with coupling strength $J_1'$ and detuning $\Delta_1$. For the qutrit $A_N$ with $N$ being odd~(even), the transition between $|e\rangle\leftrightarrow|R\rangle~(|L\rangle$) is coupled off-resonantly to resonator $B_{N-1}$ with the coupling strength $J_2'$ and detuning $\Delta_2$. In addition, an auxiliary classical field is introduced to drive the transition off-resonantly between $|e\rangle \leftrightarrow |P\rangle$ for qutrit $A_1$ with Rabi frequency $\Omega_1$ and detuning $\Delta_1$. The other classical field with Rabi frequency $\Omega_2$ and detuning $\Delta_2$ drives the transition of $A_N$ off-resonantly, $|e\rangle \leftrightarrow |L\rangle~(|R\rangle$), with $N$ being odd~(even). The interaction Hamiltonians for the first and last qutrits coupled to their adjacent resonators can be written as
\begin{subequations}
\begin{equation}\label{e9a}
H_{1}=J_1'b_1|e\rangle_{1}\langle L|e^{i\Delta_{1}t}+\Omega_1|e\rangle_{1}\langle P|e^{i\Delta_{1}t}+\rm{H.c.},
\end{equation}
\begin{equation}\label{e9b}
H_{N}=J_2'b_{N-1}|e\rangle_{N}\langle m|e^{i\Delta_{2}t}+\Omega_2|e\rangle_{N}\langle n|e^{i\Delta_{2}t}+\rm{H.c.},
\end{equation}
\end{subequations}
with $m=R~(L)$ and $n=L~(R)$ when $N$ is  odd (even), while the interaction Hamiltonian for other cells is identical with Eq.~(\ref{e1}).

\begin{figure*}
	\centering
	\includegraphics[width=0.7\linewidth]{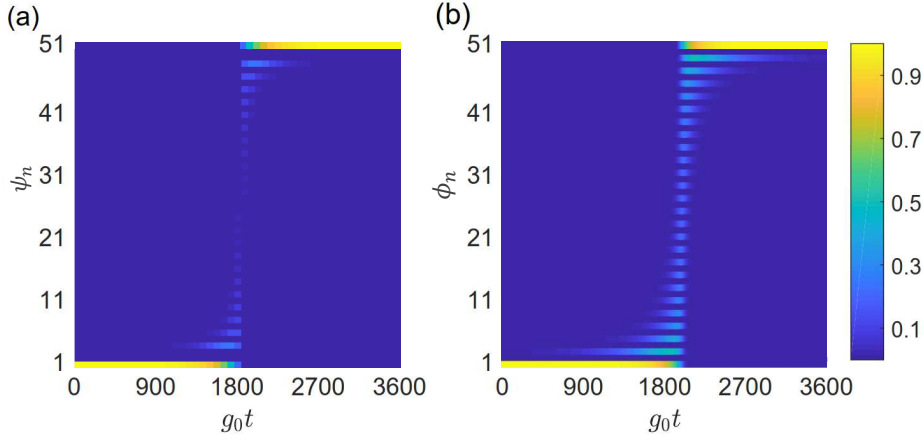}
	\caption{Distributions of the zero energy mode for (a)~Scheme B and (b)~Scheme C, respectively, on component states $|\psi_n\rangle$ and $|\phi_n\rangle$. The size of the superconducting qutrit-resonator chain is $L=2N-1=49$.}\label{f7}
\end{figure*}

For suppressing losses of qutrits in the excited state $|e\rangle$, the value of $\Delta_{1(2)}$ is supposed to be as large as possible to satisfy the condition $\Delta_{1(2)}\gg\{J_{1(2)}', \Omega_{1(2)}\}$. According to the theory of second-order perturbation~\cite{James2007}, one can eliminate adiabatically the excited states of qutrits $A_1$ and $A_N$. Besides, off-resonant-interaction-induced ground-state Stark shifts can be offset by introducing auxiliary off-resonant fields~\cite{Zhao2017}, phase compensations~\cite{Vepsaaineneaau5999}, or detuning compensations~\cite{Han:20}. Then, the Hamiltonians in Eqs.~(\ref{e9a}) and (\ref{e9b}) can be reduced into, respectively
\begin{subequations}
\begin{equation}\label{e10a}
H_{1\rm{eff}}=J_{1\rm{eff}}(|P\rangle_{1}\langle L|+|L\rangle_{1}\langle P|),
\end{equation}
\begin{equation}\label{e10b}
H_{N\rm{eff}}=J_{2\rm{eff}}(|R\rangle_{N}\langle L|+|L\rangle_{N}\langle R|),
\end{equation}
\end{subequations}
with $J_{1(2)\rm{eff}}=J_{1(2)}'\Omega_{1(2)}/\Delta_{1(2)}$, which involve solely long-lived states. When the chain with the size $L=2N-1=49$ is initially in the left edge state $|l_{\rm B}\rangle=|\psi_1\rangle=|PLR\cdots LR\rangle_{A_{25}}\otimes|000\cdots00\rangle_{B_{24}}$ under the interaction Hamiltonians (\ref{e9a}) and (\ref{e9b}), the system evolves in the finite space $\{|\psi_n\rangle\}$
\begin{equation}\label{e05}
\begin{split}
&|\psi_1\rangle=|PLR\cdots LR\rangle_{A_{25}}\otimes|000\cdots0\rangle_{B_{24}},\\
&|\psi_2\rangle=|eLR\cdots LR\rangle_{A_{25}}\otimes|000\cdots 0\rangle_{B_{24}},\\
&|\psi_3\rangle=|LLR\cdots LR\rangle_{A_{25}}\otimes|100\cdots 0\rangle_{B_{24}},\\
&|\psi_4\rangle=|LeR\cdots LR\rangle_{A_{25}}\otimes|000\cdots 0\rangle_{B_{24}},\\
&\quad\vdots\\
&|\psi_{49}\rangle=|LRL\cdots RR\rangle_{A_{25}}\otimes|000\cdots 1\rangle_{B_{24}},\\
&|\psi_{50}\rangle=|LRL\cdots Re\rangle_{A_{25}}\otimes|000\cdots 0\rangle_{B_{24}},\\
&|\psi_{51}\rangle=|LRL\cdots RL\rangle_{A_{25}}\otimes|000\cdots 0\rangle_{B_{24}}.
\end{split}
\end{equation}
To further evaluate the topological properties of zero energy states in the Scheme B, we plot the distribution of the zero mode on component states $|\psi_n\rangle$, as shown in Fig.~\ref{f7}(a). The components $|\psi_1\rangle$ and $|\psi_{51}\rangle$, which  play roles of left and right edge states, $|l_{\rm B}\rangle$ and $|r_{\rm B}\rangle$, respectively, are populated with the maximal distributions in the regions of $g_0t \in [0,1800)$ and $g_0t \in(2200,4000]$. In comparison between Figs.~\ref{f2}(d) and \ref{f7}(a), the systematic degree of freedom in the evolution subspace changes from $2N-1$ to $2N+1$ due to the introductions of classical drives on the two extremity qutrits of the chain. In the Scheme B, an extra long-lived state $|P\rangle$ is introduced in the qutrit $A_1$, and the large detuning condition is satisfied so as to eliminate adiabatically the excited state $|e\rangle$. In other words, the state transformations of Scheme A and Scheme B are, respectively
\begin{eqnarray*}
&|l_{\rm A}\rangle=|eLR\cdots R\rangle_{A_{25}}\otimes|000\cdots0\rangle_{B_{24}}  \nonumber\\
&\Downarrow\nonumber\\
&|r_{\rm A}\rangle=|LRL\cdots e\rangle_{A_{25}}\otimes|000\cdots0\rangle_{B_{24}},
\end{eqnarray*}
and
\begin{eqnarray*}
&|l_{\rm B}\rangle=|PLR\cdots LR\rangle_{A_{25}}\otimes|000\cdots0\rangle_{B_{24}}
\nonumber\\
&\Downarrow\nonumber\\
&|r_{\rm B}\rangle=|LRL\cdots RL\rangle_{A_{25}}\otimes|000\cdots 0\rangle_{B_{24}}.
\end{eqnarray*}
Thus, the distribution and spacing of bright and dark fringes in Fig.~\ref{f7}(a) are different from that in Fig.~\ref{f2}(d). The dark fringes in Fig.~\ref{f7}(a) appearing on $|\psi_2\rangle$ and $|\psi_{50}\rangle$ indicate that the population elimination of the excited state $|e\rangle$ in qutrits $A_1$ and $A_{25}$.

For sake of generating large-scale GHZ states, the superconducting qutrit-resonator chain is assumed initially in the state 
\begin{equation}\label{e11}
|\Phi'_0\rangle=\frac{1}{\sqrt{2}}\Big(|G\rangle+|l'_{\rm B}\rangle\Big).
\end{equation}
where $|l'_B\rangle=|PLR\cdots m\rangle_{A_{N}}\otimes|000\cdots0\rangle_{B_{N-1}}$ and $m=R~(L)$ when $N$ is odd~(even). After the evolution along the topologically protected zero-energy mode, similar to the Scheme A, one can obtain the final state 
\begin{equation}\label{e12}
|\Phi'_{ideal}\rangle=\frac{1}{\sqrt{2}}\Big(|G\rangle-(-1)^{N}|r'_B\rangle\Big).
\end{equation}
in which $|r'_B\rangle=|LRL\cdots n\rangle_{A_{N}}\otimes|000\cdots 0\rangle_{B_{N-1}}$ with $n=L~(R)$ when $N$ is odd~(even). Here $|R\rangle$ is used to carry the logical state 1 while $|L\rangle$ carries the logical state 0 for all qutrits. Accordingly, the initial state without considering the zero-photon resonators after the evolution becomes 
\begin{equation}\label{e7}
\frac{1}{\sqrt{2}}\left(|101\cdots 1(0)\rangle_{A_{N}}-(-1)^{N}|010 \cdots 0(1)\rangle_{A_{N}}\right).
\end{equation}
which is exactly an $N$-body GHZ state.

\begin{figure*}[htb]
\centering
\includegraphics[width=0.7\linewidth]{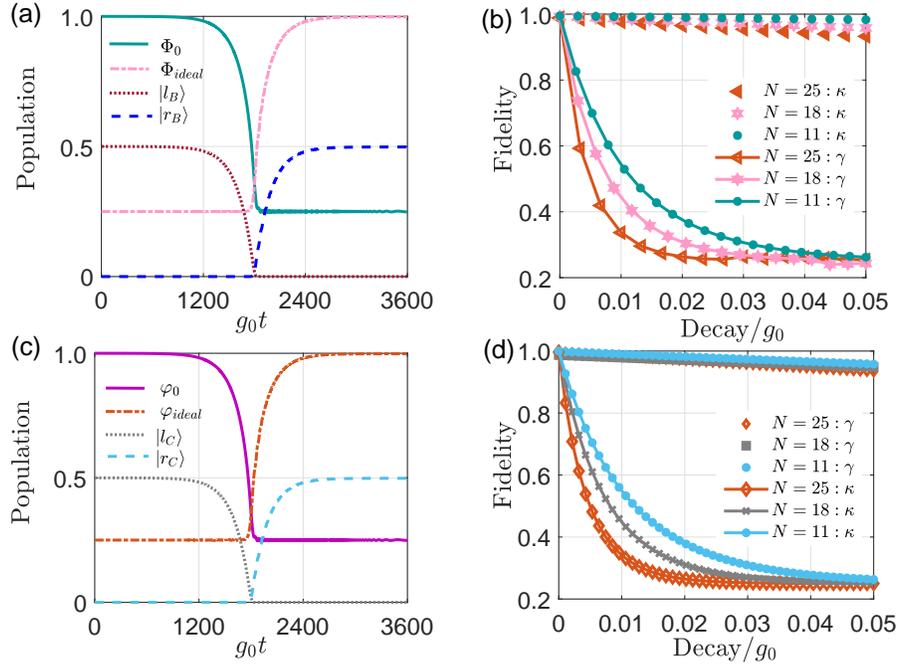}
\caption{(a) and (c) for Scheme B and Scheme C, respectively: Time evolution of population for $|\Phi_0\rangle$, $|\Phi_{ideal}\rangle$, $|\varphi_0\rangle$, $|\varphi_{ideal}\rangle$, edge states $|l_{\rm B(C)}\rangle$ and $|r_{B(C)}\rangle$ for the size of superconducting qutrit-resonator chain $L=2N-1=49$. (b) and (d) for Scheme B and Scheme C, respectively: Fidelity for the GHZ states versus the decay rate $\gamma$ or $\kappa$ of the superconducting qutrit-resonator chain for different number of qutrits, where $\gamma$~($\kappa$) in the legend represents decoherence involving only the qutrit~(resonator) decay. Parameters for (a) and (b): $J'_{1(2)}=20J_{1(2)}$, $\Delta_{12}=400g_0$, and $\Omega_{1(2)}=20g_0$.}\label{f8}
\end{figure*}

In Fig.~\ref{f8}(a), with the same parameters as in Fig.~\ref{f3} we numerically plot the time evolution of populations for the ideal state $|\Phi_{ideal}\rangle=\frac{1}{\sqrt{2}}(|RLR\cdots R\rangle_{A_{25}}+|LRL\cdots L\rangle_{A_{25}})\otimes|000\cdots0\rangle_{B_{24}}$, the initial state $|\Phi_{0}\rangle=\frac{1}{\sqrt{2}}(|RLR\cdots R\rangle_{A_{25}}+|PLR\cdots R\rangle_{A_{25}})\otimes|000\cdots0\rangle_{B_{24}}$, and two edge states $|l_{\rm B}\rangle$ and $|r_{\rm B}\rangle.$ Obviously, the population of $|\Phi_{ideal}\rangle$ ($|\Phi_0\rangle$) reaches nearly 1 (0.25) and keeps steady at the end of evolution time, which proves the feasibility of Scheme B. The populations of two edge states have the identical trend with Fig.~\ref{f3}. 

In order to verify the effectiveness of suppressing the losses of qutrits in the excited state, we simulate the fidelity of the ideal GHZ state $|\Phi_{ideal}\rangle$ by solving the non-Hermitian Liouville equation. In Fig.~\ref{f8}(b), the fidelities for 11-, 18-, and 25-body GHZ states hold on $98.3\%$, $95.8\%$, and $93.2\%$ even though $\kappa/g_0=0.05$. In addition, it is evident that the fidelity in Fig.~\ref{f8}(b) can reach values similar to that without taking account of edge states in Fig.~\ref{f5}. When $\gamma/g_0=0.01$, the fidelities of 11-, 18-, and 25-body GHZ states maintain $53.2\%$, $45.2\%$, and $33.6\%$, respectively. With regard to $\gamma$, the fidelity of the ideal state decreases more slightly and displays a more robust result than Scheme A because of the suppression of excitation populations in qutrits $A_1$ and $A_N$ during the whole evolution.

\subsection{Scheme C for suppressing excitation of all qutrits}
As shown in Figs.~\ref{f6}(c) and (d), for the qutrit $A_1$, the transition between $|e\rangle \leftrightarrow |L\rangle$ is coupled resonantly to the resonator $B_1$ with coupling strength $J_{1}$. When $N$ is odd (even), the transition between $|e\rangle\leftrightarrow|R\rangle~(|L\rangle$) in the last qutrit $A_N$ is coupled resonantly to the resonator $B_{N-1}$ with the coupling strength $J_{2}$. Also, one classical field drives the transition resonantly between $|e\rangle \leftrightarrow |P\rangle$ for qutrit $A_1$ with Rabi frequency $\tilde{\Omega}_{1}$. Under the condition of $N$ being odd (even), the other classical field with Rabi frequency $\tilde{\Omega}_{2}$ drives resonantly the transition $|e\rangle \leftrightarrow |L\rangle ~(|R\rangle)$ in qutrit $A_N$. Therefore, the interaction Hamiltonian involving the first and last qutrits can be written as 
\begin{equation}\label{e13}
\begin{split}
H_{1'}&=J_{1}b_1|e\rangle_{1}\langle L|+\tilde{\Omega}_{1}|e\rangle_{1}\langle P|+\rm{H.c.},\\
H_{N'}&=J_{2}b_{N-1}|e\rangle_{N}\langle m|+\tilde{\Omega}_{2}|e\rangle_{N}\langle n|+\rm{H.c.},
\end{split}
\end{equation}
where $m=R~(L)$ and $n=L~(R)$ when $N$ is odd~(even).  And the interaction Hamiltonian of other cells still keeps consistent with Eq.~(\ref{e1}). The shapes of the coupling strengths are engineered as Gauss functions
\begin{equation*}
\begin{split}
\tilde{\Omega}_{1}=J_2=g_0\exp{[-(t-3\tau)^2/\tau^2]},\\
\tilde{\Omega}_{2}=J_1=g_0\exp{[-(t-2\tau)^2/\tau^2]},
\end{split}
\end{equation*}
where $J_1$ and $J_2$ are reverse with respect to Eq.~(\ref{e5}).

We take $N=25$ as an example, and when the chain is initially in the left edge state $|l_{\rm C}\rangle=|\phi_1\rangle=|PLR\cdots R\rangle_{A_{25}}\otimes|000\cdots0\rangle_{B_{24}}$. The system evolves in the finite subspace $\{|\phi_n\rangle\}$
\begin{equation}\label{e14}
\begin{split}
&|\phi_1\rangle=|PLR\cdots R\rangle_{A_{25}}\otimes|000\cdots0\rangle_{B_{24}},\\
&|\phi_2\rangle=|eLR\cdots R\rangle_{A_{25}}\otimes|000\cdots0\rangle_{B_{24}},\\
&|\phi_3\rangle=|LLR\cdots R\rangle_{A_{25}}\otimes|100\cdots0\rangle_{B_{24}},\\
&|\phi_4\rangle=|LeR\cdots R\rangle_{A_{25}}\otimes|000\cdots0\rangle_{B_{24}},\\
&\quad\vdots\\
&|\phi_{49}\rangle=|LRL\cdots R\rangle_{A_{25}}\otimes|000\cdots1\rangle_{B_{24}},\\
&|\phi_{50}\rangle=|LRL\cdots e\rangle_{A_{25}}\otimes|000\cdots0\rangle_{B_{24}},\\
&|\phi_{51}\rangle=|LRL\cdots R\rangle_{A_{25}}\otimes|000\cdots0\rangle_{B_{24}}.
\end{split}
\end{equation}
Then we plot the distribution of the zero energy mode on component states $|\phi_n\rangle$ in Fig.~\ref{f7}(b). Distinctly, both the two ends of evolutionary states $|\phi_1\rangle$ and $|\phi_{51}\rangle$ are populated with the maximal distributions in the regions of $g_0t \in [0,1800)$ and $g_0t\in(2200,4000]$, respectively. In comparison between Figs.~\ref{f2}(d) and \ref{f7}(b), the system degree of freedom in the evolution subspace changes from $2N-1$ to $2N+1$ due to the introductions of classical drives on the two extremity qutrits of the chain. In Fig.~\ref{f7}, the distribution and spacing of bright and dark fringes are different from that of Scheme A or Scheme B. The bright fringes in Scheme A or Scheme B are distributed in the excitations of qutrits, while in Scheme C they are distributed in the excitations of resonators. Therefore, in Scheme C excitations of all qutrits are suppressed.

So as to obtain a large-scale GHZ states, the initial state of the chain is prepared in Eq.~(\ref{e11}). The first state component $|G\rangle$ does not evolve, while the second state component $|l'_{B}\rangle$ evolves into $(-1)^{N}|r'_{B}\rangle$. Hence, the final state is
\begin{eqnarray}
\begin{split}\label{e15}
|\Psi_F\rangle=&\frac{1}{\sqrt{2}}\left(|G\rangle+(-1)^{N}|r'_{B}\rangle\right).
\end{split}
\end{eqnarray}
which has a minus sign difference from Eq.~(\ref{e12}). In Fig.~\ref{f8}(c), we plot the time evolution of populations for the initial state $|\varphi_{0}\rangle=|\Phi_0\rangle$ , the ideal state $|\varphi_{ideal}\rangle=\frac{1}{\sqrt{2}}(|RLR\cdots R\rangle_{A_{25}}-|LRL\cdots L\rangle_{A_{25}})\otimes|000\cdots0\rangle_{B_{24}}$, two edge states $|l_{C}\rangle=|l'_{B}\rangle$ and $|r_{\rm C}\rangle=|r'_{B}\rangle$ for generating a 25-body GHZ state. As expected, the population of $|\varphi_{ideal}\rangle$ ~($|\varphi_{0}\rangle$) is close to unity (0.25) at the time $g_0t=3600$ and remains stable. As for the two edge states, the population in Scheme C has the same climate as that in Scheme A and Scheme B. The result reveals that the above theoretical analysis is correct and feasible. Figure \ref{f8}(d) shows the relationship between the fidelity and the decay rate $\gamma$ or $\kappa$. As the decay rates increase, the fidelity of the ideal GHZ state decreases. As opposed to the loss mechanisms of Scheme A and Scheme B, the decay rate $\kappa$ of resonators has a greater influence on the fidelity. When $\kappa/g_0=0.01$, the fidelity is reduced to $56.1\%$ for the 11-body GHZ state. In contrast among the six lines, the fidelity is relatively immune to the decay of qutrits. The fidelity can retain at $95.9\%$, $95.0\%$, and $93.6\%$ with $\gamma/g_0=0.05$, which indicates that Scheme C displays the most robust performance to resist the losses of qutrits in the superconducting qutrit-resonator chain among the three schemes.

In Scheme C, the fidelity of the ideal large-scale GHZ state is affected least by the losses of qutrits, compared with Scheme A and Scheme B. However, the cost is that the losses of resonators have a greatest influence on the fidelity, which indicates that the Scheme C requires resonators to hold a long coherence time. While according to the interaction of Scheme A and Scheme B, it is required that the qutrits have a long coherence time, but without a strict requirement on the quality factor of resonators. Thus, corresponding to the performance of experimental devices in the superconducting qutrit-resonator chain, we can choose different schemes to realize large-scale GHZ states with high fidelity in the case of loss mechanisms by adjusting the energy level structure of qutrits and coupling strengths.

\begin{figure}
\centering
\includegraphics[width=0.95\linewidth]{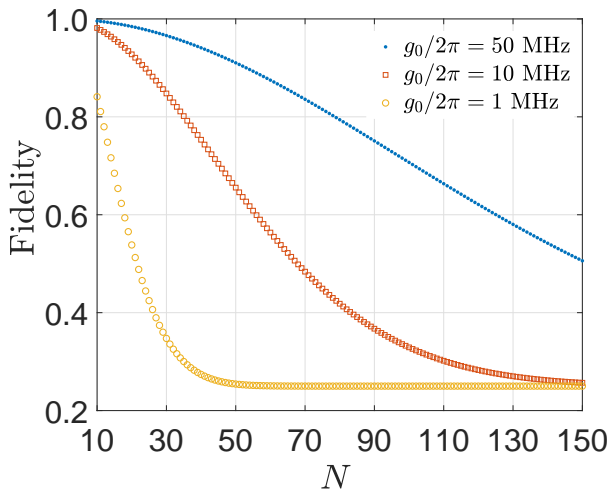}
\caption{Fidelity of the GHZ state with the scale of entanglement, $N$, under the feasible coupling strengths (e.g., $g_0/2\pi=50$ MHz, 10 MHz and 1 MHz) and  coherence times of qutrits and resonators in Scheme A. We choose coherence time of the qutrits and resonators as $\tau_a=\tau_b=1$~ms.}\label{f9}
\end{figure}
\section{Scale of GHZ states}\label{S5}
In experiment, by designing a $\pi$-phase difference across the Josephson junction in circuit to restrain the energy relaxation induced by quasiparticle dissipation, one can obtain a qutrit with coherence time over 1~ms~\cite{Pop2014}. As for a resonator, the coherence time of the photons in the resonator can be much longer~\cite{GU2017}. Up to now, the superconducting resonator lifetimes between 1~ms and 10~ms have been reported~\cite{Reagor2013,Reagor2016,Axline2016}.  Generally, the typical feasible coupling strength can be modulated in the range of 1 MHz to 50 MHz~\cite{Mundada2019}, providing a considerable adjustability in experiment. 

In this section, we discuss an accessible GHZ state scale $N$ by taking into account the experimentally available coherence times of qutrits and resonators under the condition of feasible coupling strengths. For convenience, we take Scheme A as an example. Figure \ref{f9} shows the relationship between fidelities of GHZ states and the scale of entanglement $N$ with feasible $g_0$ and coherence times of qutrits ($\tau_a=1/\gamma=1$~ms) and resonators $\tau_b=1/\kappa=1$~ms). Obviously, the fidelity exhibits the decreasing tendency with increasing $N$. Regarding to feasible coupling strengths, the fidelity declines at different rates. Under the condition of $g_0/2\pi=50$~MHz, fidelities of 10- to 50-body GHZ states keep above $90\%$. Even if the scale of a GHZ state is $N=150$, its fidelity would still reach $50.60\%$. As for $g_0/2\pi=10$~MHz, the fidelity of 67-body GHZ state stays above $50\%$. The fidelity can be over $90\%$ with $N$ less than 23. However, the GHZ state fidelity has the fastest rate of decline in the case of $g_0/2\pi=1$~MHz, where the fidelity of a GHZ state can achieve $50\%$  with the maximum entanglement scale $N=20$. Under the existing experimental conditions, the higher fidelity of large-scale GHZ states can be generated with the higher value of coupling strengths.

\section{Experimental consideration and prospective improvement}\label{S6}
\subsection{Device and initial state}
Benefiting from the rapid development in circuit-QED technologies, the circuit-QED system provides us an excellent experimental platform to realize large-scale GHZ states proposed in our work. We can construct a circuit-QED system via arranging the transmission line resonator and the superconducting qutrits in the space, as shown in Fig.~\ref{f1}(b). According to the existing circuit-QED technology, the qubit-resonator chain systems with on the order of 10–20 qubits have been demonstrated~\cite{Kelly2015,Kandala2017,Neill195,Otterbach2017,Verifying2020,Mooney2019}. Simultaneously, a chain of 72 superconducting resonators coupled via transmons can be realized  ~\cite{Fitzpatrick2017}. The scale of qutrit-resonator chain manufacturing in a metal chip is generally from micron to millimeter~\cite{Majer2009,Hybrid2020,Circuit2010,Quantum2020}. In general, the flux qutrit is a superconducting circuit made up of Josephson junctions or/and capacitance~\cite{Manucharyan113}. While the flux qutrit can be operated at any applied external flux through the flux qutrit loop. The resonator is composed by a linear inductance in parallel with a capacitance, which can be fabricated from a NbN film deposited on a sapphire substrate~\cite{Fast2020}. The coupler is replaced by two Josephson junctions with a SQUID loop to realize tunable coupling: the magnetic flux that threads this loop determines its effective inductance $E_{J}(\phi)\sim E_{J}(0)\cos(2\pi\phi/\Phi_0)$~\cite{Garc2020}. Josephson junctions can be designed by the metallic chip (Al/AlO$_{x}$/Al) using the electron-beam lithography~\cite{Junction2011,Circuit2010}. In experiment, the circuit-QED device is usually operated in a single-shot liquid $^3$He with a base millikelvin temperatures and heavily filtered cryogenic microwave lines~\cite{Majer2009,Quantum2020}.

In the above schemes, specified initial states are needed to prepare in Eqs.~(\ref{e6}) and (\ref{e11}) to realize $N$-body GHZ states. These specified initial states denote $A_1$ in the superposition $(|R\rangle+|e(P)\rangle)/\sqrt{2}$, $B_1$ in $|0\rangle$, $A_2$ in $|L\rangle$, $\cdots$, $A_N$ in $|R\rangle$ ($|L\rangle$) when $N$ is odd (even). In experiment, the preparation of the initial states can be conducted through two steps: (i) As reported recently in Ref.~\cite{Driving2018}, the qutrit holding a $\Lambda$-type structure can be cooled to one of its two lowest energy eigenstates ($|R\rangle$ or $|L\rangle$) by resonator decay through spontaneous Raman scattering so as to obtain the state $|RLR\cdots R(L)\rangle_{A_{N}}$. (ii) A $\pi/2$ pulse is applied to the qutrit $A_1$ to create a superposition state of $(|R\rangle + |e(P)\rangle)/\sqrt{2}$~\cite{Experimental2019}. The resonator keeps in the zero photon Fock state. Finally, the qutrit-resonator chain can be prepared in $1/\sqrt{2}(|RLR\cdots R(L)\rangle_{A_N}+|P(e)LR\cdots R(L)\rangle_{A_N})\otimes|000\cdots0\rangle_{B_{N-1}}$. 

\subsection{Tunable couplings}
In order to achieve high-fidelity GHZ states, tunable couplings among qutrits and resonators are required. Tunable couplers of both varieties have been realized in several experiments: between two tunable qubits~\cite{Chen2014,Harrabi2009}, between a qubit and a resonator~\cite{Allman2014,Srinivasan2011}, and between resonators~\cite{Wang2011,Zakka2011}. Particularly, a direct tunable coupler is realized by a tunable circuit element between the qutrit and the resonator, e.g., a flux-biased direct-current SQUID to generate strong resonant and nonresonant tunable interactions between a qubit and a lumped-element resonator~\cite{Allman2014}.

In this work, we adopt a direct tunable coupler replaced by SQUID between qutrits and resonators. The coupling strengths $J_1$ and $J_2$ can be tunable through adopting controlled voltage pulses generated by an AWG to tune the flux threading the SQUID loop~\cite{Majer2009}. For example, the resonator $B_{N}$ is coupled resonantly to transitions $|e\rangle \leftrightarrow |m\rangle$ ($m \in \{R,L\}$). The coupling strength $J_{1(2)}$ can be expressed by~\cite{Possible2003,Quantum2004}
\begin{equation}\label{e16}	
J_{1(2)}=\frac{1}{L}\sqrt{\frac{\omega_{B_{N}}}{2\mu_{0}\hbar}}\langle e|\Phi|m\rangle \int_{S} \textbf{B}_{B_{N}}(\vec{r},t)\cdot d\textbf{S},
\end{equation}
where $\textbf{S}$ is the surface bounded by the loop of SQUID, $\omega_{B_{N}}$ the resonator frequencies of $B_{N}$. Accordingly, $\textbf{B}_{B_{N}}(\vec{r},t)$ is the the magnetic components of the $B_{N}$ resonator mode. For a standingwave resonators, $\textbf{B}_{B_{N}}(\vec{r},t)=\mu_{0}\sqrt{2/V_N}\cos k_{N}z_{N}$ ($k_N$ is the wave number of $B_{N}$ resonator, $V_N$ and $z_N$ are the $B_{N}$ resonator volume and the $B_{N}$ resonator anxis). In this case, a modulating field can be added when a qutrit works on its optimal frequency point~\cite{Reagoreaao3603}, and thus does not cause the shortening of qutrit coherent times.

\subsection{Accelerating adiabatic process}
Over the past decade, techniques of shortcuts to adiabaticity~(STA)~\cite{Chen2010,Odelin2019} receive a lot of attention, because STA can accelerate adiabatic processes but remain the robustness of adiabatic processes. Recently, a fast quantum state transfer from the left edge state to the right edge state in a topological SSH chain with next-to-nearest-neighbor~(NNN) interaction was presented, which provides the simplest one-dimensional lattice with protected edge state~\cite{Felippo2020}. The idea of this approach is based on an engineering of NNN interactions between the sites of the chain, which exactly cancels nonadiabatic couplings due to an imperfect condition of adiabatic evolution. Inspired by counterdiabatic driving methods~\cite{Berry2009,Demirplak2003}, the quantum chain without limitation of the adiabaticity constraint, combined with a dynamical control of NNN interactions, is thus governed by the following Hamiltonian $H(t)$:
\begin{eqnarray}\label{e17}	
H(t)&=&H_0(t)+H_c(t),\nonumber\\
H_0(t)&=&\sum_{n=1}^{N-1}t_2(t)|B_n\rangle \langle A_n|+t_1(t)|A_{n+1}\rangle \langle B_n|+\rm{H.c.},\nonumber\\
H_c(t)&=&\sum_{n=1}^{N-1}i\alpha_{n}(t)|A_{n+1}\rangle \langle A_{n}|+\rm{H.c.},
\end{eqnarray}
where $H_c(t)$ is the control Hamiltonian literally canceling the nonadiabatic couplings. The time-dependent NNN hoppings $\alpha_n(t)$ is only to cancel nonadiabatic transitions from the time-dependent eigenvector $|\phi_0(t)\rangle=\mathcal{N}\sum^{N}_{n=1}(\frac{-t_2}{t_1})^{n-1}|A_n\rangle$, where $\mathcal{N}$ is the normalization constant.

Analogously, with the help of such a method for speeding up the adiabatic state transfer in the SSH model, it is possible to find a control Hamiltonian, added to the initial Hamiltonian, that literally cancels the nonadiabatic couplings in the topological model. Therefore, it is of great potential to realize fast preparation of a large-scale GHZ state by engineering the NNN interaction between qutrits, which may constrain the impact of systematic decoherence and thus enhance the fidelity of generating large-scale GHZ states.

In addition, an alternative method, Floquet-engineering STA~(FESTA)~\cite{Petiziol2018,Claeys2019}, may also be applied in speeding up the generation of a large-scale GHZ state. FESTA is an effective succedaneum of counterdiabatic driving methods, which does not require an additional control Hamiltonian but is combined with a periodic driving component to oscillate the initial Hamiltonian~\cite{Petiziol2018}. This Floquet-engineering periodic driving component will form an effective connection between transferred two states so as to offset exactly nonadiabatic couplings~\cite{Petiziol2020}. FESTA has been demonstrated for quantum state transfer in spin chains in a recent experiment~\cite{Peng2019SB}. Therefore, it is of great potential to use FESTA in our model for accelerating the adiabatic generation of GHZ states by replacing the adiabatic pulses with the Floquet-engineering oscillating pulses.

\section{Conclusion}\label{S7}
In conclusion, we have proposed a model of a superconducting qutrit-resonator chain, and the topological edge state with zero energy is analytically derived. Along this topological zero-energy mode, a state transfer from one extremity qutrit to the other of the chain can be implemented, accompanied with state flips of all intermediate qutrits. Three schemes are shown for generating large-scale GHZ states that are protected by the topological zero-energy mode and thus hold great robustness against disorder of the qutrit-resonator coupling strengths. The effect of losses induced by decay of qutrits and resonators on entanglement fidelity is investigated, and the results indicate that the three schemes meet different performance requirements of experimental devices. Furthermore, we study the accessible entanglement scale when taking into account the experimentally available coherence times of qutrits and resonators, and find that with the maximum of the qutrit-resonator coupling strengths $g_0/2\pi=50$~MHz, it is possible to achieve a $50$-body GHZ state with the fidelity $F>0.9$, and a $150$-body GHZ state with $F>0.5$. Finally, we discuss the experimental consideration of generating GHZ states, including the physical implementation of the model, preparation of the initial states, and tunable couplings in circuit-QED chain, and also show the potential to accelerate adiabatic process of generating GHZ states. The present work is expected to be helpful for promoting the experimental improvement of large-scale GHZ states.
\section*{ACKNOWLEDGEMENTS}
This work was supported by National Natural Science Foundation of China (NSFC) (Grant No.11675046), Program for Innovation Research of Science in
Harbin Institute of Technology (Grant No. A201412), and Postdoctoral Scientific Research Developmental Fund of Heilongjiang Province (Grant No. LBH-Q15060).
\bibliography{apssamp}
\end{document}